\documentclass[aps,physrev,reprint,superscriptaddress,twocolumn]{revtex4-2}

\usepackage{graphicx}
\usepackage{hyperref}
\usepackage{physics}
\usepackage{bm}

\begin{document}


\title{Parameter optimization for the unimon qubit}


\author{Rostislav Duda}
\email[]{rostislav.duda@aalto.fi}
\affiliation{QCD Labs, QTF Centre of Excellence, Department of Applied Physics, Aalto University, Aalto 00076, Finland}

\author{Eric Hyyppä}
\affiliation{IQM Quantum Computers, Keilaranta 19, Espoo 02150, Finland}

\author{Olli Mukkula}
\affiliation{CSC--IT Center for Science Ltd., Keilaranta 14, Espoo 02150, Finland}

\author{Vasilii Vadimov}
\affiliation{QCD Labs, QTF Centre of Excellence, Department of Applied Physics, Aalto University, Aalto 00076, Finland}

\author{Mikko Möttönen}
\affiliation{QCD Labs, QTF Centre of Excellence, Department of Applied Physics, Aalto University, Aalto 00076, Finland}
\affiliation{VTT Technical Research Centre of Finland Ltd. $\&$ QTF Centre of Excellence, P.O. Box 1000, 02044 VTT, Finland}


\date{\today}

\begin{abstract}
Inductively shunted superconducting qubits, such as the unimon qubit, combine high anharmonicity with protection from low-frequency charge noise, positioning them as promising candidates for the implementation of fault-tolerant superconducting quantum computers. 
In this work, we develop accurate closed-form approximations for the frequency and anharmonicity of the unimon qubit that are also applicable to any single-mode superconducting qubits with a single-well potential profile, such as the quarton qubit or the kinemon qubit. We use these results to theoretically explore the single-qubit gate fidelity and coherence times across the parameter space of qubits with a single-well potential. We find that the gate fidelity can be optimized by tuning the Hamiltonian to (i) a high qubit mode impedance of 1--2 k$\Omega$, (ii) a low qubit frequency of 1 GHz, (iii) and a perfect cancellation of the linear inductance and the Josephson inductance attained at a flux bias of half flux quantum. 
According to our theoretical analysis, the proposed qubit parameters have potential to enhance the single-qubit gate fidelity of the unimon beyond 99.99\% even without significant improvements to the dielectric quality factor or the flux noise density measured for the first unimon qubits.
Furthermore, we compare unimon, transmon and fluxonium qubits in terms of their energy spectra and qubit coherence subject to dielectric loss and $1/f$ flux noise in order to highlight the advantages and limitations of each qubit type.
\end{abstract}

\maketitle

\section{Introduction}
\label{sec:introduction}
Despite its brief history~\cite{Feynman1982}, quantum computing has made significant progress in delivering on the promise of computational advantage in practical applications~\cite{Arute2019, Zhong2020, Wu2021, Daley2022, King2024}. However, the search for an optimal qubit is still ongoing. Over the past two decades, multiple competing qubit modalities have emerged, such as trapped ions~\cite{Cirac1995}, quantum dots~\cite{Loss1998}, optical photons~\cite{Knill2001}, neutral atoms~\cite{Bluvstein2024}, and superconducting microwave circuits~\cite{Devoret2004} which gave rise to the research field of circuit quantum electrodynamics (cQED)~\cite{Krantz2019, Kjaergaard2020, Blais2021}. Within each modality, various qubit types are being developed for optimized fidelity of quantum algorithm execution.

Superconducting qubits have come a long way since the introduction of the Cooper-pair box \cite{Bouchiat1998}, and currently, transmon qubits \cite{Koch2007}, consisting of a Josephson junction (JJ) shunted by a large capacitance, are the most utilized for building multi-qubit processors. The large capacitive shunt of the transmon provides exponential protection against low-frequency charge noise at the expense of reduced anharmonicity. The state-of-the-art transmons can reach millisecond coherence times~\cite{tuokkola2024methods}, single-qubit gate infidelity below $10^{-4}$~\cite{Li2023} and two-qubit gate infidelity of $10^{-3}$~\cite{Li2024}, although experimental realization of a large-scale chip that concurrently possesses all of these properties has not yet been reported in the literature. Furthermore, typical anharmonicities of the transmon qubit are relatively low, of the order of 5\% of the qubit frequency~\cite{Barends2013}, limiting the speed of qubit operations and qubit readout.

Introduction of an inductive shunt for a JJ opens new pathways for enhancing the qubit anharmonicity and the fidelity of the quantum gates. In fluxonium qubits, the inductance of the shunt greatly exceeds the Josephson inductance, leading to a double-well potential energy at the half-flux-quantum sweet spot~\cite{Manucharyan2009}. 
Fluxonia have recently achieved millisecond coherence times while approaching single-qubit gate infidelities of $10^{-5}$~\cite{rower2024suppressing}.  
In the parameter regime of fluxonium qubits, it is experimentally feasible to reach a high anharmonicity exceeding the qubit frequency by an order of magnitude. However, high-coherence fluxonia typically exhibit a low qubit frequency of a few hundred megahertz or below \cite{zhang2021universal, somoroff2023millisecond, rower2024suppressing}, which may lead to high thermal population and limit the speed of operations 
\cite{Ding2023_FTF, rower2024suppressing, zwanenburg2025single}. 
Furthermore, practical implementation of fluxonium qubits requires the use of superinductors, which are typically realized with Josephson junction arrays~\cite{Masluk2012, Ranni2023}. This additional layer of fabricational complexity may pose a challenge for reproducibility of such qubits in large-scale quantum processors. 

In the recently introduced unimon qubit~\cite{Hyyppa2022}, the inductive energy of the shunt and the Josephson energy are of similar magnitude, resulting in a single-well potential and a partial or full cancellation of the quadratic potential-energy terms at a flux bias of half flux quantum. 
As a result, the quartic potential energy dominates at the bottom of the potential well, leading to an enhanced anharmonicity of the qubit. The gradiometric structure of the realized unimons~\cite{Hyyppa2022} provides a partial protection against flux noise. In principle, the multimode structure of the unimon qubit may also enable hosting multiple qubits in a single unimon circuit, enhancing the potential for scalability~\cite{Tuohino2024}. The first experiments with the unimon qubit demonstrated anharmonicities as high as 20\% of the qubit frequency~\cite{Hyyppa2022}. However, significant improvements are needed to reach the state-of-the-art performance of transmons and fluxonia, since the measured coherence times of the first devices were of the order of 10 $\mu$s, with the single-qubit gate infidelity reaching $10^{-3}$~\cite{Hyyppa2022}.

In this work, we explore the parameter space of the unimon qubit using newly introduced analytical techniques combined with numerical studies. Subsequently, we find design parameters that maximize the single-qubit gate fidelity and coherence time. Importantly, we find a parameter regime, where we predict a single-qubit gate fidelity above 99.99\% even without significant improvements to the dielectric quality factor or flux noise density of the previously realized unimon qubits~\cite{Hyyppa2022}. Furthermore, our comparison of the optimized unimon with transmons and fluxonia reveals that the the unimon is a serious contester for the existing state-of-the-art qubit types.

The remainder of the paper is organized as follows: In Sec.~\ref{sec:variational_method}, we develop accurate closed-form approximations for the qubit frequency and anharmonicity that are valid for any single-mode superconducting qubit with a single-well potential. In Sec.~\ref{sec: average gate infidelity estimation}, we outline our approach for estimating the average single-qubit gate infidelity of the unimon qubit. In Sec.~\ref{sec:results}, we study the average gate infidelity and coherence times as functions of the qubit parameters, and compare the energy spectra and coherence properties of unimons, transmons, and fluxonia. We present our conclusions in Sec.~\ref{sec:conclusions}.

\section{Variational approximations of eigenstates and eigenenergies}
\label{sec:variational_method}

\begin{figure*}
    \includegraphics[width=0.95\textwidth]{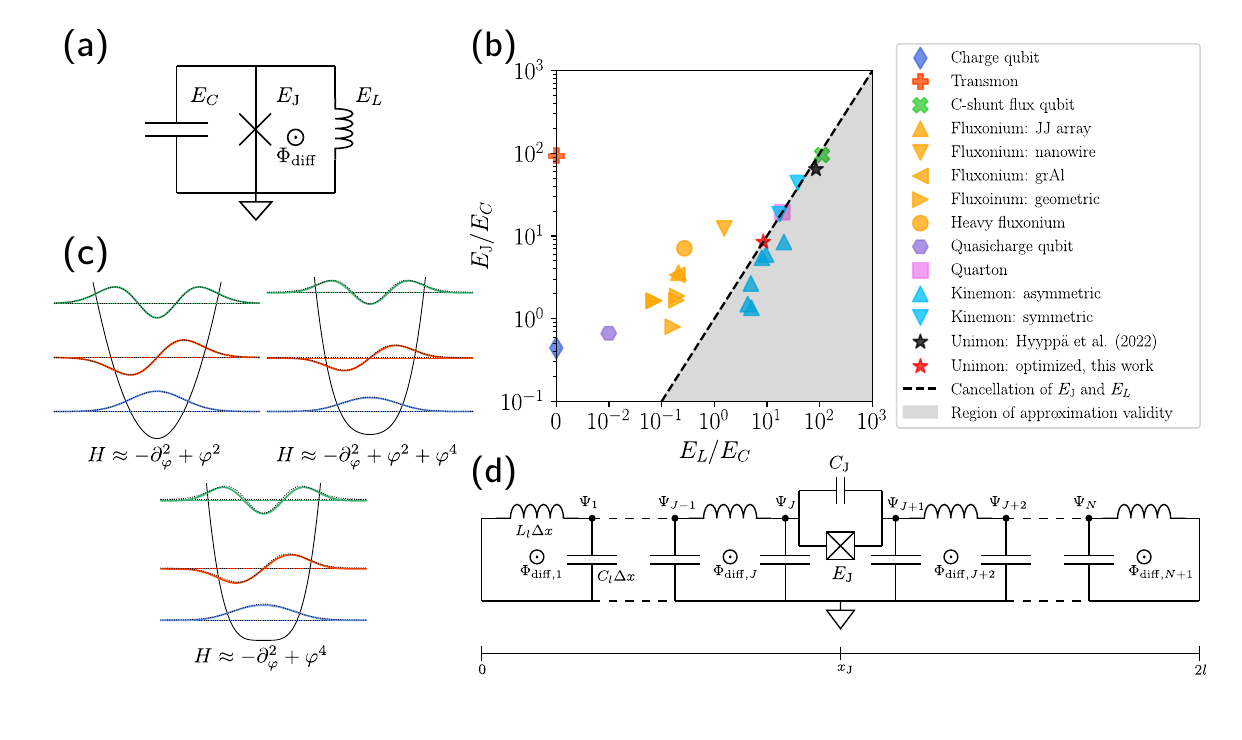}
    \caption{(a) Lumped-element circuit model of the unimon qubit corresponding to the Hamiltonian of Eq.~(\ref{eq:H_0}). (b) Parameter landscape of superconducting circuits, including the region of validity of the proposed approximation scheme (gray shading)~\cite{Nakamura1999, Barends2013, Yan2016, Manucharyan2009, Hazard2019, Rieger2023, peruzzo2021geometric, Earnest2018, Pechenezhskiy2020, Yan2020, Kalacheva2023, Hyyppa2022}. The black star denotes the experimentally characterized parameters of the first unimon qubits in Ref.~\citep{Hyyppa2022}, whereas the red star denotes the proposed optimized parameters based on the theoretical analysis of Sec.~\ref{sec: Gate fidelity optimization}. (c) Energy landscape for a few representative instances of the potential, where numerically obtained eigenstates and eigenenergies are shown in solid color and the almost overlapping analytically obtained approximate results are denoted by dashed lines. (d) Distributed-element circuit model of the unimon qubit.}
        \label{fig:0}
\end{figure*}

To gain insight into the energy spectrum of the unimon without relying on numerical methods, we analytically derive closed-form approximations for the three lowest eigenstates of the unimon circuit presented in Fig.~\ref{fig:0}. Our derivation begins with the single-mode model of the unimon~\cite{Hyyppa2022}. In this approximation, we express the Hamiltonian of the qubit mode as
\begin{equation}
\label{eq:H_0}
    \hat{H}_{0} = 4E_{C}\hat{n}^{2} + \frac{1}{2}E_{L}\hat{\varphi}^{2} - E_{\mathrm{J}} \cos{\left(\hat{\varphi} - \varphi_{\mathrm{diff}} \right)},
\end{equation}
where $\left[\hat{\varphi}, \hat{n}\right] = \mathrm{i}$, $\varphi_{\mathrm{diff}} = 2\pi \Phi_\mathrm{diff} / \Phi_0$ is the phase bias provided by the external flux $\Phi_\mathrm{diff}$, $\Phi_0 = \pi \hbar / e$ is the flux quantum, $\hbar$ is the reduced Planck constant, $e$ is the elementary charge, $E_{C}$ is the effective capacitive energy of the qubit mode, $E_{L}$ is the effective inductive energy of the qubit mode, and $E_{\mathrm{J}}$ is the Josephson energy of the junction. Below, we work in the phase basis $\{|\varphi\rangle\}$, in terms of which we may express $\langle \varphi'|\hat{\varphi}|\varphi\rangle = \delta(\varphi-\varphi')\varphi$ and $\langle \varphi'|\hat{n}|\varphi\rangle = \delta(\varphi-\varphi')(-\mathrm{i}\partial_{\varphi})$. At the half flux quantum sweet spot $\varphi_{\mathrm{diff}} = \pi$, where the anharmonicity of the qubit is maximized, the relevant Hamiltonian in the phase basis can be simplified into
\begin{equation}
    {H}_{0} = -4E_{C}\partial^{2}_{\varphi} + \frac{1}{2}E_{L}\varphi^{2} + E_{\mathrm{J}} \cos{\varphi}.
\end{equation}
Subsequently, we rescale ${H}_{0}$ by the capacitive energy and define ${H}_{1} \equiv {H}_{0}/(4E_{C})$:
\begin{equation}
\label{eq:H_dimless_full}
    {H}_{1} = -\partial^{2}_{\varphi} + \frac{\varepsilon_{2} + \varepsilon_{4}}{2}\varphi^{2} + \varepsilon_{4} \cos{\varphi},
\end{equation}
where $\varepsilon_{2} = (E_{L} - E_{\mathrm{J}})/(4E_{C})$ and $\varepsilon_{4} = E_{\mathrm{J}}/(4E_{C})$ are dimensionless parameters corresponding to quadratic and quartic coefficients in a Taylor expansion of the potential energy. Next, we truncate ${H}_{1}$ up to the fourth order in $\varphi$ using the Maclaurin series of $\cos \varphi$, which yields
\begin{equation}
\label{eq:H_dimless_taylor}
    {H}_{2} = -\partial^{2}_{\varphi} + \frac{1}{2}\varepsilon_{2}\varphi^{2} + \frac{1}{24}\varepsilon_{4}\varphi^{4}.
\end{equation}
We restrict our analytical study to the single-well regime, i.e., we work in the region where $\varepsilon_{2} \geq 0$, $\varepsilon_{4} \geq 0$, excluding the case $\varepsilon_{2}=\varepsilon_{4}=0$. In particular, $\varepsilon_{2} \geq 0$ implies that $E_{L} \geq E_{\mathrm{J}}$, an operation regime, that is relevant not only for the unimon, but also for other recently introduced qubits, such as quarton \cite{Ye2024}, kinemon \cite{Kalacheva2023}, and C-shunt flux qubits under the single-mode approximation \cite{Yan2016}. Thus, our results are applicable beyond unimon qubits.

Naïvely, one may attempt to employ the first-order perturbation theory to obtain estimates for the eigenenergies by treating the quartic potential energy term as a perturbation to a harmonic oscillator. However, this approach produces energy correction terms of the form $E_{n}^{(1)} \propto \varepsilon_{4}/\varepsilon_{2}$, which diverge as $\varepsilon_{2}$ tends to zero. Consequently, we need a different strategy to handle the scenario where the quartic contribution to the potential energy becomes dominant. To this end, we resort to the variational method to find approximate closed-form eigenfunctions corresponding to the three lowest eigenenergies of ${H}_{2}$ \cite{Sakurai2020}. This method can be summarized as follows: given some system with a Hamiltonian $\hat{H}$, we approximate its $N$ lowest eigenstates with a choice of ansatz states $\{\ket{\psi_{n}(\pmb{\theta}_{n})}\}_{n=0}^{N-1}$, where $\pmb{\theta}_{n}$ is a vector of parameters for the $n$-th eigenstate. These parameters are determined through the energy minimization criterion in conjunction with orthogonality constraints as
\begin{align}
\label{eq:var_crit}
    \nabla_{\!\pmb{\theta}_{n}}\!\!&\expval{\hat{H}}{\psi_{n}(\pmb{\theta}_{n})} = \mathbf{0}, \\
    &\braket{\psi_{m}(\pmb{\theta}_{n})}{\psi_{n}(\pmb{\theta}_{n})} = \delta_{mn}.
\end{align}
In practice, a suitable choice of ansatz states is the deciding factor that contributes to the accuracy of the results. 

For our purposes, we choose the eigenfunctions inspired by the quantum harmonic oscillator as our ansatz functions. The oscillator eigenfunctions are composed of Hermite polynomials weighted by Gaussian envelopes of fixed and identical widths. We decide to treat the width of each Gaussian and the relevant coefficients in the polynomials as the optimization parameters for the variational method. 
This choice leads us to the following ansatz functions
\begin{align}
    \label{eq:psi0_ansatz}
    \psi_{0}(\varphi; \theta_{0}) &= \left(\frac{\theta_{0}}{\pi}\right)^{1/4} \mathrm{e}^{-\frac{1}{2}\theta_{0}\varphi^{2}}, \\
    \label{eq:psi1_ansatz}
    \psi_{1}(\varphi; \theta_{1}) &= \left(\frac{\theta_{1}}{\pi}\right)^{1/4}\sqrt{2\theta_{1}}\,\varphi \, \mathrm{e}^{-\frac{1}{2}\theta_{1}\varphi^{2}}, \\
    \label{eq:psi2_init}
    \psi_{2}(\varphi; \begin{bmatrix}
        \theta_{2} & \lambda
    \end{bmatrix}) &= \left(\frac{\theta_{2}}{\pi}\right)^{1/4}\frac{2\theta_{2}\left(\lambda\varphi^{2}-1\right)\mathrm{e}^{-\frac{1}{2}\theta_{2}\varphi^{2}}}{\sqrt{3\lambda^{2} - 4\lambda\theta_{2} + 4\theta_{2}^{2}}},
\end{align}
where $\{\theta_{n}\}_{n=0}^{2}$ correspond to the widths of the Gaussian envelopes and $\lambda$ is a parameter that we determine by imposing the following orthogonality relation between $\psi_{0}$ and $\psi_{2}$:
\begin{equation}
\label{eq:orth_cond}
    \int\limits_{-\infty}^{\infty} \psi_{2}^{*}(\varphi; \begin{bmatrix}
        \theta_{2} & \lambda
    \end{bmatrix})\,\psi_{0}(\varphi; \theta_{0})\,\mathrm{d}\varphi = 0.
\end{equation}
Note that for any choice of the parameters, the orthogonality of $\psi_{n}$ and $\psi_{n+1}$, $n\in \{0,1\}$ is guaranteed by their even and odd symmetries with respect to the reflection $\varphi \to -\varphi$. Evaluation of Eq.~\eqref{eq:orth_cond} leads to a simple orthogonality condition
\begin{equation}
    \lambda = \theta_{0} + \theta_{2},
\end{equation}
turning Eq.~\eqref{eq:psi2_init} into
\begin{equation}
    \label{eq:psi2_ansatz}
    \psi_{2}(\varphi; \theta_{2}) = \left(\frac{\theta_{2}}{\pi}\right)^{1/4}\frac{2\theta_{2}\left[\left(\theta_{0} + \theta_{2}\right)\varphi^{2}-1\right]}{\sqrt{3\theta_{0}^{2} + 2\theta_{0}\theta_{2} + 3\theta_{2}^{2}}} \, \mathrm{e}^{-\frac{1}{2}\theta_{2}\varphi^{2}},
\end{equation}
where $\theta_{0}$ does not appear as an argument of the wavefunction since it is effectively a fixed parameter determined through the minimization of the ground-state energy. We then apply the variational principle to the ansatz functions of Eqs.~\eqref{eq:psi0_ansatz}, \eqref{eq:psi1_ansatz}, and \eqref{eq:psi2_ansatz}, which calls for us to solve
\begin{equation}
\label{eq:var_crit_2}
    \frac{\partial}{\partial\theta_{n}}\int\limits_{-\infty}^{\infty}\psi_{n}^{*}(\varphi; \theta_{n}) \, {H}_{2} \,\psi_{n}(\varphi; \theta_{n})\,\mathrm{d}\varphi = 0
\end{equation}
for $n \in \{0, 1, 2\}$. All of these integrals can be evaluated analytically, leading to the following concise energy minimization conditions for parameters $\theta_{0}$ and $\theta_{1}$:
\begin{align}
    \theta_{0}^{3} - \frac{1}{2}\varepsilon_{2}\theta_{0} - \frac{3}{24}\varepsilon_{4} = 0, \\
    \theta_{1}^{3} - \frac{1}{2}\varepsilon_{2}\theta_{1} - \frac{5}{24}\varepsilon_{4} = 0.
\end{align}
These are depressed cubic equations that admit closed-form solutions~\cite{Zwillinger2002}. The situation becomes more involved with $\theta_{2}$, as the additional orthogonality constraint on $\psi_{2}$ gives rise to a septic polynomial equation
\begin{equation}
\label{eq:theta_2_eq_full}
\begin{aligned}
    &\theta_{2}^{3}\underbrace{\left(\theta_{0}+3\theta_{2}\right)\left(7\theta_{0}^{3}+15\theta_{0}^{2}\theta_{2}+5\theta_{0}\theta_{2}^{2}+5\theta_{2}^{3}\right)}_{p_{0}} \\
    &-\frac{1}{2}\varepsilon_{2}\theta_{2}\underbrace{\left(3\theta_{0}+\theta_{2}\right)\left(5\theta_{0}^{3}+5\theta_{0}^{2}\theta_{2}+15\theta_{0}\theta_{2}^{2}+7\theta_{2}^{3}\right)}_{p_{1}} \\
    &-\frac{1}{24}\varepsilon_{4}\underbrace{\left(105\theta_{0}^{4}+180\theta_{0}^{3}\theta_{2}+310\theta_{0}^{2}\theta_{2}^{2}+244\theta_{0}\theta_{2}^{3}+57\theta_{2}^{4}\right)}_{p_{2}}\\
    &=0,
\end{aligned}
\end{equation}
which is not analytically solvable in general~\cite{Abel1826}. Through numerical evaluation of the roots of the above equations, we find that the inequality $1 \leq \theta_{2}/\theta_{0}<2$ holds across a vast parameter range of  $10^{-6}\leq \varepsilon_{2},\varepsilon_{4} \leq10^{6}$. In this $\theta_{2}/\theta_{0}$ region, the ratios of the polynomial factors $p_{0}$, $p_{1}$, and $p_{2}$ are approximately constant, taking values of $p_{1}/p_{0}\approx 1$, and $p_{2}/p_{0} \approx 7$. Consequently, through division of Eq.~\eqref{eq:theta_2_eq_full} by $p_{0}$, one can approximate Eq.~\eqref{eq:theta_2_eq_full} with a cubic equation
\begin{equation}
    \theta_{2}^{3} - \frac{1}{2}\varepsilon_{2}\theta_{2} - \frac{7}{24}\varepsilon_{4} = 0.
\end{equation}
Each of the resulting cubic equations admits three solutions, and hence an additional set of constraints is necessary to ensure that an appropriate root is chosen. In particular, we require $\theta_{n}$ to be real and positive in order for the wavefunctions to be normalizable. The relevant roots of these three equations in different $\varepsilon_{2}$ regimes can be expressed in compact forms given by
\begin{equation}
\label{eq:ansatz_root_0}
    \theta_{0} = \begin{cases}
        \frac{\sqrt{6\varepsilon_{2}}}{3}\cos\left[\frac{1}{3}\cos^{-1}\left(\frac{3\sqrt{6}}{8}\frac{\varepsilon_{4}}{\varepsilon_{2}^{3/2}}\right)\right] & \text{if} \quad \frac{3\sqrt{6}}{8}\frac{\varepsilon_{4}}{\varepsilon_{2}^{3/2}} \leq 1 \\
        \frac{\sqrt{6\varepsilon_{2}}}{3}\cosh\left[\frac{1}{3}\cosh^{-1}\left(\frac{3\sqrt{6}}{8}\frac{\varepsilon_{4}}{\varepsilon_{2}^{3/2}}\right)\right] & \text{if} \quad \frac{3\sqrt{6}}{8}\frac{\varepsilon_{4}}{\varepsilon_{2}^{3/2}} \geq 1 \\
        \frac{1}{2}\sqrt[3]{\varepsilon_{4}} & \text{if} \quad \varepsilon_{2}=0,
    \end{cases}
\end{equation}
\begin{equation}
\label{eq:ansatz_root_1}
    \theta_{1} = \begin{cases}
        \frac{\sqrt{6\varepsilon_{2}}}{3}\cos\left[\frac{1}{3}\cos^{-1}\left(\frac{5\sqrt{6}}{8}\frac{\varepsilon_{4}}{\varepsilon_{2}^{3/2}}\right)\right] & \text{if} \quad \frac{5\sqrt{6}}{8}\frac{\varepsilon_{4}}{\varepsilon_{2}^{3/2}} \leq 1 \\
        \frac{\sqrt{6\varepsilon_{2}}}{3}\cosh\left[\frac{1}{3}\cosh^{-1}\left(\frac{5\sqrt{6}}{8}\frac{\varepsilon_{4}}{\varepsilon_{2}^{3/2}}\right)\right] & \text{if} \quad \frac{5\sqrt{6}}{8}\frac{\varepsilon_{4}}{\varepsilon_{2}^{3/2}} \geq 1 \\
        \frac{1}{2}\sqrt[3]{\frac{5\varepsilon_{4}}{3}} & \text{if} \quad \varepsilon_{2}=0, \\
    \end{cases}
\end{equation}
and
\begin{equation}
\label{eq:ansatz_root_2}
    \theta_{2} = \begin{cases}
        \frac{\sqrt{6\varepsilon_{2}}}{3}\cos\left[\frac{1}{3}\cos^{-1}\left(\frac{7\sqrt{6}}{8}\frac{\varepsilon_{4}}{\varepsilon_{2}^{3/2}}\right)\right] & \text{if} \quad \frac{7\sqrt{6}}{8}\frac{\varepsilon_{4}}{\varepsilon_{2}^{3/2}} \leq 1 \\
        \frac{\sqrt{6\varepsilon_{2}}}{3}\cosh\left[\frac{1}{3}\cosh^{-1}\left(\frac{7\sqrt{6}}{8}\frac{\varepsilon_{4}}{\varepsilon_{2}^{3/2}}\right)\right] & \text{if} \quad \frac{7\sqrt{6}}{8}\frac{\varepsilon_{4}}{\varepsilon_{2}^{3/2}} \geq 1 \\
        \frac{1}{2}\sqrt[3]{\frac{7\varepsilon_{4}}{3}} & \text{if} \quad \varepsilon_{2}=0.
    \end{cases}
\end{equation}

Equations \eqref{eq:psi0_ansatz}, \eqref{eq:psi1_ansatz}, and \eqref{eq:psi2_ansatz} together with Eqs.~\eqref{eq:ansatz_root_0}--\eqref{eq:ansatz_root_2} form the desired set of approximate eigenfunctions. Conveniently, we recover the exact eigensolutions of the quantum harmonic oscillator in the limit $\varepsilon_{4} \rightarrow 0$. 

To find the analytical expressions for the eigenenergies, we compute the mean of the Hamiltonian $H_1$ with respect to these eigenfunctions as $E_{n}=4E_{C}\int\psi_{n}^{*}{H}_{1}\psi_{n}\mathrm{d}\varphi$, obtaining
\begin{widetext}
\begin{align}
    E_{0}&=2\theta_{0}E_{C} + \frac{1}{4\theta_{0}}E_{L} + \mathrm{e}^{-1/(4\theta_{0})}E_{\mathrm{J}}, \label{eq: E0 variational approx}\\
    E_{1}&=6\theta_{1}E_{C} + \frac{3}{4\theta_{1}}E_{L} + \left(1-\frac{1}{2\theta_{1}}\right)\mathrm{e}^{-1/(4\theta_{1})}E_{\mathrm{J}}, \label{eq: E1 variational approx} \\
    E_{2} &=
    2A\theta_{2}E_{C} + \frac{B}{4\theta_{2}}E_{L}+\left(1-\frac{C}{\theta_{2}}+\frac{D}{\theta_{2}^{2}}\right)\mathrm{e}^{-1/(4\theta_{2})}E_{\mathrm{J}}, \label{eq: E2 variational approx}
\end{align}
\end{widetext}
where the coefficients $A,B,C,$ and $D$ are given by
\begin{align}
    A &= \frac{7\theta_{0}^{2}+18\theta_{0}\theta_{2}+15\theta_{2}^{2}}{3\theta_{0}^{2}+2\theta_{0}\theta_{2}+3\theta_{2}^{2}}\approx 5, \\
    B &= \frac{15\theta_{0}^{2}+18\theta_{0}\theta_{2}+7\theta_{2}^{2}}{3\theta_{0}^{2}+2\theta_{0}\theta_{2}+3\theta_{2}^{2}} \approx 5,\\
    C &= \frac{3\theta_{0}^{2}+4\theta_{0}\theta_{2}+\theta_{2}^{2}}{3\theta_{0}^{2}+2\theta_{0}\theta_{2}+3\theta_{2}^{2}} \approx 1,\\
    D &= \frac{\theta_{0}^{2}+2\theta_{0}\theta_{2}+\theta_{2}^{2}}{12\theta_{0}^{2}+8\theta_{0}\theta_{2}+12\theta_{2}^{2}} \approx \frac{1}{8}.
\end{align}
Using Eqs.~\eqref{eq: E0 variational approx}--\eqref{eq: E2 variational approx}, the qubit frequency $f_{01}$ and the anharmonicity $\alpha/(2\pi)$ are further obtained as $$f_{01}=\omega_{01}/(2\pi) = (E_1 - E_0)/h,$$ and $$\alpha/(2\pi) = (E_2 - 2E_1 + E_0)/h,$$ respectively. Here, we emphasize, that we use the ansatz functions obtained from the variational principle to the Hamiltonian $H_2$, but use the Hamiltonian $H_1$ to obtain the energy levels. This can be justified in the spirit of the first-order perturbation theory provided that the fourth-order expansion well approximates the potential energy in $H_1$ on the scale of $|\varphi| \lesssim \theta_n^{-1/2}$, $n\in\{0,1,2\}$. In the single-well regime, this assumption is valid for the lowest energy levels of the qubit.

\begin{figure*}
    \includegraphics[width=\linewidth]{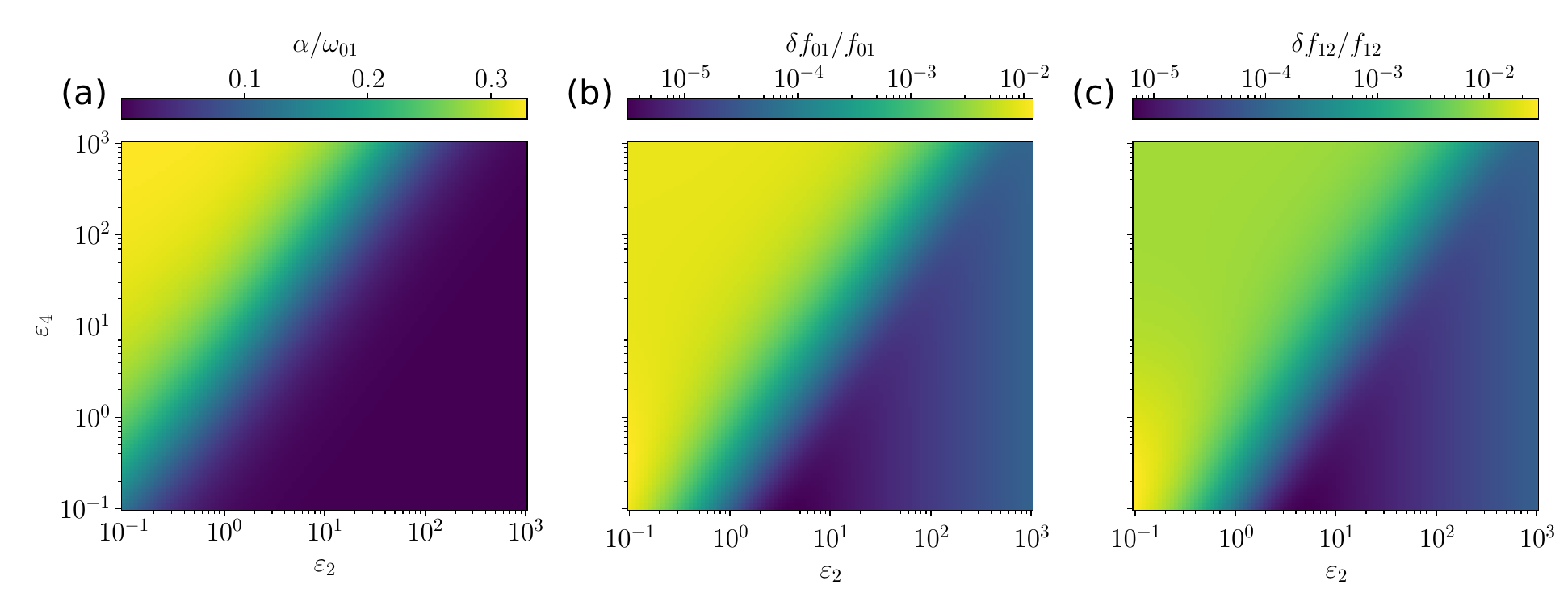}
    \caption{(a) Relative anharmonicity computed with the variational ansatz wavefunctions as a function of parameters $\varepsilon_{2}$ and $\varepsilon_{4}$ of the Hamiltonian ${H}_{1}$. Subfigures (b) and (c) show relative error of the variational estimates for the eigenstate transition frequencies $f_{01}=(E_{1}-E_{0})/h$ and $f_{12}=(E_{2}-E_{1})/h$ with respect to the numerically obtained values.}
    \label{fig:1}
\end{figure*}

The variational approximations introduced above provide accurate eigenenergy estimates for ${H}_{1}$ in a wide range of parameter values $\varepsilon_{2}$ and $\varepsilon_{4}$, as illustrated in Fig.~\ref{fig:1}. Importantly, this approach agrees with numerical simulations in a region that spans five orders of magnitude of $\varepsilon_{2}$ and $\varepsilon_{4}$ with relative errors of the order of at most $\sim\! 1\%$ for the transition frequencies $f_{01}$ and $f_{12}$, combined with a maximum error of 4\% for the resulting anharmonicity. The relative anharmonicity $\alpha/\omega_{01}$ appears to have an upper bound of approximately $1/3$, obtained when the potential of ${H}_{1}$ becomes effectively quartic to the lowest order, i.e., for $\varepsilon_{4} \gg \varepsilon_{2}$. This observation is in line with previously conducted numerical studies on the quarton qubit~\cite{Yan2020}. The variational method may also be used with higher-order Taylor expansions of Eq.~(\ref{eq:H_dimless_full}). However, the fourth-order solution that we employ strikes an appealing balance between accuracy and simplicity.

\section{Average gate infidelity estimation}
\label{sec: average gate infidelity estimation}

Qubit design is a broad and challenging task which can be formulated at its core as an optimization problem. In the context of the unimon qubit, our task is to find the energy parameters $E_{C}$, $E_{L}$, and $E_{\mathrm{J}}$ that minimize the average gate infidelity owing to decoherence \cite{Pedersen2007}
\begin{equation}
    \bar{\mathcal{E}} = 1 - \frac{1}{6}\left\{3 + \mathrm{e}^{-\Gamma_{1}t_{\mathrm{g}}} + 2\mathrm{e}^{-\frac{\Gamma_{1}}{2}t_{\mathrm{g}}}\mathrm{Re}\left[f(t_{\mathrm{g}})\right]\right\}, \label{eq: infidelity}
\end{equation}
where $t_{\mathrm{g}}$ is the single-qubit gate duration, and $\Gamma_{1}$ is the longitudinal relaxation rate or decay rate. The function $f(t)$ describes a nonexponential decay envelope that accounts for pure dephasing due to the quadratically coupled $1/f$ flux noise source \cite{Ithier2005} at the sweet spot of the unimon. To solve this optimization problem, we need to express the longitudinal relaxation rate $\Gamma_1$, the gate duration $t_{\mathrm{g}}$, and the contribution of pure dephasing to decoherence $f(t_{\mathrm{g}})$ in terms of the unimon parameters $E_{C}$, $E_{L}$, and $E_{\mathrm{J}}$. 

According to the first experiments with the unimon qubit~\cite{Hyyppa2022}, the dominant relaxation mechanisms of the qubit are $1/f$ flux noise and dielectric losses. In our calculations, we decompose $\Gamma_{1}$ into a sum of the rates corresponding to these two mechanisms as
\begin{equation}
\label{eq:gamma_1}
    \Gamma_{1} = \Gamma_{1}^{1/f} + \Gamma_{1}^{\mathrm{diel}}.
\end{equation}
The relaxation rates $\Gamma_{1}^{1/f}$ and $\Gamma_{1}^{\mathrm{diel}}$ are further given by \cite{Hyyppa2022}
\begin{align}
\label{eq:gamma_1overf}
    \Gamma_{1}^{1/f} &= \mu E_{L}^{2} \omega_{01}^{-1} \lvert \langle 0 \lvert \hat{\varphi} \rvert 1 \rangle \rvert^{2}, \\
\label{eq:gamma_die}
    \Gamma_{1}^{\mathrm{diel}} &= \eta E_{C} \lvert \langle 0 \lvert \hat{n} \rvert 1 \rangle \rvert^{2} \coth\left(\frac{\hbar \omega_{01}}{2k_{\mathrm{B}}T}\right),
\end{align}
where $T$ is the temperature of the dielectric medium, $k_{\mathrm{B}}$ is the Boltzmann constant, and $\{\lvert n \rangle\}_{n=0}^{\infty}$ are the eigenstates of $\hat{H}_{0}$. In addition, we have introduced two free parameters, $\mu$ and $\eta$. The parameter $\mu$ is related to the flux noise power spectral density (PSD) at one hertz $A_{\Phi_{\mathrm{diff}}}$ through
\begin{equation}
\label{eq:param_mu}
    \mu = \frac{8\pi^{3}}{\hbar^{2}} \frac{A_{\Phi_{\mathrm{diff}}}^{2}}{\Phi_{0}^{2}},
\end{equation}
and parameter $\eta$ is in turn related to the dielectric quality factor $Q_{\mathrm{diel}}$, or conversely the dielectric loss tangent $\tan\delta_{\mathrm{diel}} = 1/Q_{\mathrm{diel}}$ by
\begin{equation}
\label{eq:param_eta}
    \eta = \frac{16}{\hbar Q_{\mathrm{diel}}} = \frac{16}{\hbar}\tan\delta_{\mathrm{diel}}.
\end{equation}
Our calculations assume that the flux noise PSD and the dielectric loss factor remain constant as we vary the Hamiltonian, unless stated otherwise. In practice, changes to the Hamiltonian can be realized by varying the geometry of the qubit, which may either increase or decrease $A_{\Phi_{\mathrm{diff}}}$~\cite{Braumuller2020}. Similarly, $Q_{\mathrm{diel}}$ may be enhanced through materials engineering and geometric optimizations \cite{Lahtinen2020}. However, investigation of the effects of qubit geometry on operation fidelity is more suited for a case-by-case study, and is outside of the scope of this work. The parameters $\mu$ and $\eta$ are chosen based on the experimental data of the first unimon qubit experiments~\cite{Hyyppa2022} presented in Table~\ref{tab:params}.

Next, we express the single-qubit gate duration $t_{\mathrm{g}}$ in terms of the qubit parameters. In general, decoherence dominates the gate error at long gate durations, whereas leakage to non-computational states yields most of the gate error at short durations if the qubit frequency significantly exceeds the anharmonicity. Thus, the optimal gate duration is tightly connected to the qubit anharmonicity, in particular, the two are inversely related to each other \cite{Mortzoi2009}
\begin{equation}
    t_{\mathrm{g}} = \frac{2\pi\nu}{|\alpha|}, \label{eq: speed limit}
\end{equation}
where the parameter $\nu$ encodes the effect of the employed control pulse on the optimal gate duration. 
Neglecting potential limitations of control electronics and crosstalk, the current theoretical lower bound for the duration of the single-qubit gate is reached around $\nu \sim 1$ \cite{Theis2016, Zhu2021}. Modern state-of-the-art pulse shaping techniques are rapidly catching up to this bound: carefully optimized control pulses \cite{Werninghaus2021, Hyyppa2024} are able to achieve $\nu \approx 1.3$. Due to such a marginal discrepancy between the theoretically proposed bound and the latest experimental developments, we have set $\nu=1$ in this work, effectively evoking the assumption of optimal control pulse shaping.

Let us estimate the effects of pure dephasing. At the sweet spot $\varphi_\mathrm{diff} = \pi$, the $1/f$ flux noise couples quadratically to the unimon qubit. This type of coupling leads to a nonexponential decay law $f(t)$ with distinct features at long- and short-time regimes. With our choice of $t_{\mathrm{g}}$, we are interested in the short-time regime, where $f(t)$ is given by \cite{Ithier2005}
\begin{equation}
\label{eq:decay_law}
    f(t) = \frac{1}{\sqrt{1-2\mathrm{i}\kappa A_{\Phi_{\mathrm{diff}}}^{2}t\ln{\frac{1}{\omega_{\mathrm{ir}}t}}}},
\end{equation}
where $\omega_{\mathrm{ir}}$ is the infrared cutoff frequency and $\kappa= \partial^{2}\omega_{01}/\partial\Phi_{\mathrm{diff}}^{2}$ is the flux curvature of the angular frequency of the qubit. This decay law is valid for $t_{\mathrm{g}} < (\kappa A_{\Phi_{\mathrm{diff}}}^{2}/2)^{-1}$, an assumption we have found to hold in the parameter range under investigation. We introduced $\omega_{\mathrm{ir}}$ to treat the divergence of the $1/f$ spectrum at low frequencies. The choice of $\omega_{\mathrm{ir}}$ is determined by the details of the experiment, such as signal acquisition time \cite{Nakamura2002, Makhlin2004}, and in our calculations we have set $\omega_{\mathrm{ir}}$ to $1\,\mathrm{kHz}$. Due to the logarithm, the exact value of $\omega_{\mathrm{ir}}$ has only a minor effect on the resulting pure dephasing rate. We determine the flux curvature with the help of a second-order perturbative expansion \cite{Mukkula2024}
\begin{equation}
    \kappa = 4\pi^{2}\frac{E_{L}^{2}}{\hbar}\left(\sum_{m\neq1}\frac{|\langle m| \hat{\varphi}|1\rangle|^{2}}{E_{1}-E_{m}} - \sum_{n\neq 0}\frac{|\langle n|\hat{\varphi}|0\rangle|^{2}}{E_{0}-E_{n}}\right),
\end{equation}
truncated to the three lowest eigenstates. This truncation is justified since, at half-flux quantum, the phase matrix elements related to higher-energy transitions of the unimon rapidly decay and their contribution to the curvature may be neglected. 

Upon inspection of Eq.~\eqref{eq:decay_law}, we find it sensible to define the pure dephasing quasirate $\widetilde{\Gamma}_{\varphi}^{1/f}$ as
\begin{equation}
    \widetilde{\Gamma}_{\varphi}^{1/f} \equiv 2 \kappa A_{\Phi_{\mathrm{diff}}}^{2},
\end{equation}
where the tilde signifies that this quasirate corresponds to a power law decay of Eq.~\eqref{eq:decay_law}, although the notion of a fixed decay rate is ill-defined in this scenario. Note that the conventional additive relaxation rate relation does not hold in the case of a nonexponential decay mechanism. In the following analysis, we study the average gate infidelity across constrained sweeps over the space of parameters $(E_{\mathrm{J}}, E_{L}, E_{C})$, while keeping the free parameters $(\mu, \eta, \nu)$ constant.

\begin{table}[t]
    \caption{Experimentally obtained data from the highest-performing qubit (Qubit B) of Ref.~\cite{Hyyppa2022} used for our estimations of the average gate infidelity, including flux noise PSD at one hertz $A_{\Phi_{\mathrm{diff}}}$, dielectric quality factor $Q_{\mathrm{diel}}$, Josephson energy $E_{\mathrm{J}}$, inductive energy $E_{L}$, capacitive energy $E_{C}$, angular qubit frequency $\omega_{01}$, and bath temperature $T$.}
    \begin{center}
        \begin{tabular}{l|l|l|l|l|l|l}
            \hline
            \hline
            $A_{\Phi_{\mathrm{diff}}}$ & $Q_{\mathrm{diel}}$ & $E_{\mathrm{J}}/h$ & $E_{L}/h$ & $E_{C}/h$ & $\omega_{01}/(2\pi)$ & $T$ \\
            ($\mu\Phi_{0}$) & & (GHz) & (GHz) & (GHz) & (GHz) & (mK)\\ \hline
            15.0 & $3.5\times10^{5}$ & 19.0 & 25.2 & 0.297 & 4.488 & 25 \\ \hline
            \hline
        \end{tabular}
    \end{center}
    \label{tab:params}
\end{table}

\section{Results}
\label{sec:results}
\subsection{Gate fidelity optimization}
\label{sec: Gate fidelity optimization}
The methods outlined in Secs.~\ref{sec:variational_method} and \ref{sec: average gate infidelity estimation} allow us to assess the effect of the qubit design parameters on the average gate infidelity. The three-dimensional space of energies $(E_{\mathrm{J}}, E_{L}, E_{C})$ may be non-trivial to explore, and therefore, we constrain it by considering cross-sections of fixed qubit frequency $f_{01}$. Furthermore, we choose our design parameters to be the energy ratio $E_{\mathrm{J}}/E_{L}$ and the qubit mode impedance $Z$, given by
\begin{equation}
    Z = \frac{R_{\mathrm{K}}}{4\pi}\sqrt{\frac{2E_{C}}{E_{L}}}, \label{eq: impedance}
\end{equation}
where $R_{\mathrm{K}} = 2\pi \hbar / e^2$ is the von Klitzing constant. This impedance is different from the characteristic impedance $Z_{0} = \sqrt{L_{l}/C_{l}}$ of a coplanar waveguide (CPW) resonator that embeds the JJ of the unimon qubit shown in Fig.~\ref{fig:0}(d), where $L_{l}$ and $C_{l}$ are the inductance and capacitance per unit length, respectively. In particular, for a CPW geometry with a JJ in the middle, we have $Z=\sqrt{12}Z_{0}$. 

These design variables define a two-dimensional parameter surface $(E_{\mathrm{J}}/E_{L}, Z) \in [0, 1] \times [0.1\,\mathrm{k}\Omega, 5\,\mathrm{k}\Omega]$. The variational approximation developed in Sec.~\ref{sec:variational_method} is used to efficiently find the inverse mapping $(f_{01},E_{\mathrm{J}}/E_{C},Z) \rightarrow (E_{\mathrm{J}}, E_{L}, E_{C})$. Given the Hamiltonian energy parameters for the sweep, we diagonalize $\hat{H}_{0}(E_{\mathrm{J}}, E_{L}, E_{C})$ numerically to ensure the accuracy of our results. We then study anharmonicity, flux curvature, and average gate infidelity at different qubit frequencies to determine optimal values for the ratio $E_{\mathrm{J}}/E_{L}$ and the mode impedance $Z$.

\begin{figure*}
    \includegraphics[width=\linewidth]{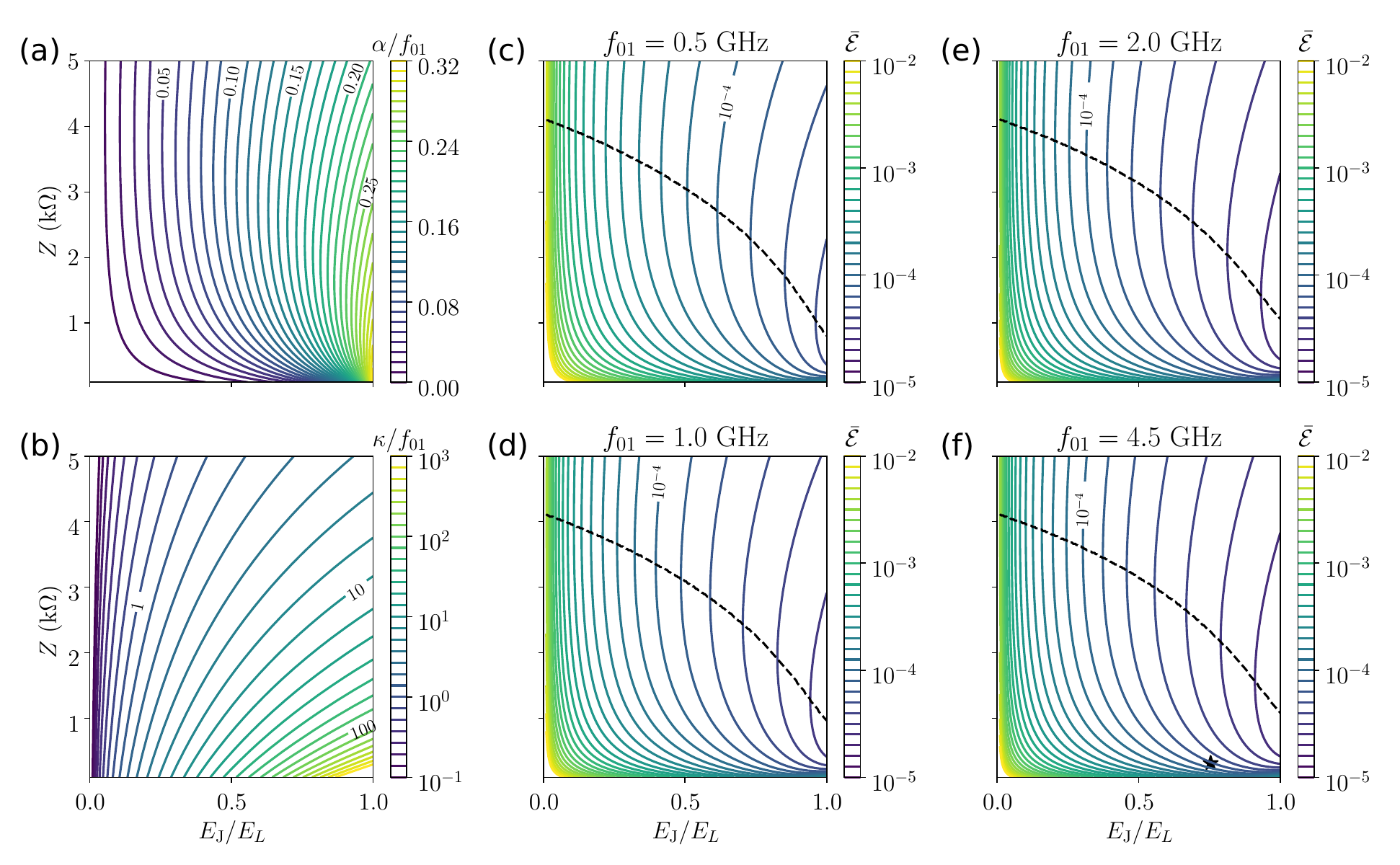}
    \caption{(a)--(b) Relative anharmonicity $\alpha/f_{01}$ and relative flux curvature $\kappa/f_{01}$ as functions of the energy ratio $E_{\mathrm{J}}/E_{L}$ and qubit mode impedance $Z$. The results in panels (a) and (b) are independent of qubit frequency. (c)--(f) As (a), but for the average gate infidelity $\bar{\mathcal{E}}$ and for a fixed qubit frequency of  (c) $f_{01}=0.5$ GHz, (d) $f_{01}=1.0$ GHz, (e) $f_{01}=2.0$ GHz, and (f) $f_{01}=4.5$ GHz. The dashed lines trace the contours of minimal gate infidelity as a function of $E_\mathrm{J}/E_L$. The black star indicates the unimon qubit parameters used in~\cite{Hyyppa2022}.}
    \label{fig:gate_fidelity_optimization}
\end{figure*}

Based on the relative anharmonicity shown in Fig.~\ref{fig:gate_fidelity_optimization}(a), we find that the maximum of $\sim\! 1/3$ is obtained when the impedance is low, and $E_{\mathrm{J}}=E_{L}$, corresponding to the perfect cancellation of the quadratic flux term in the unimon Hamiltonian. While maximizing the anharmonicity may intuitively seem favorable, the flux curvature presented in Fig.~\ref{fig:gate_fidelity_optimization}(b) increases with increasing anharmonicity, leading to enhanced dephasing. Conversely, the relative anharmonicity decreases linearly, and the relative flux curvature reduces exponentially as impedance increases, suggesting that an optimal fidelity may be attained at a high impedance.

This hypothesis is confirmed by investigating the average gate infidelity across the parameter space for different qubit frequencies ranging from $f_{01}=0.5$ GHz to $f_{01} = 4.5$ GHz as shown in Figs.~\ref{fig:gate_fidelity_optimization}(c)--(f). The contour profiles demonstrate that the minimum gate infidelity is achieved at an impedance in the range of 1--4 k$\Omega$ depending on the $E_{\mathrm{J}}/E_{L}$ ratio. Notably, $Z$ required for optimal gate infidelity does not increase without bounds and settles at an admittedly high but nonetheless experimentally realizable value~\cite{Stockklauser2017, Niepce2019, Grünhaupt2019, Ranni2023}. At high impedances, the gate infidelity is reduced as $E_{\mathrm{J}}/E_{L}$ approaches unity, and the optimum of infidelity is reached at the energy cancellation spot $E_{\mathrm{J}}=E_{L}$ when $Z\approx1\,\text{k}\Omega$.

The above-found behavior of the average gate infidelity remains largely identical across different qubit frequencies, but we observe that the infidelity $\bar{\mathcal{E}}$ slightly increases at lower qubit frequencies due to the 
temperature dependent term of the dielectric loss rate in Eq.~\eqref{eq:gamma_die}. At $f_{01}=4.5$ GHz, the estimated infidelity $\bar{\mathcal{E}}$ reaches  $2\times10^{-5}$ at the optimal point, corresponding to nearly five nines of single-qubit gate fidelity despite the relatively high dielectric loss tangent and flux noise density assumed in the calculations. However, our estimates do not account for potential limitations of control electronics on the achievable gate duration. In practice, a lower qubit frequency may provide a lower gate infidelity if the bandwidth, sampling rate, or maximum power of the drive electronics prevent reaching the optimal gate duration in Eq.~\eqref{eq: speed limit}.   Overall, our findings in this section suggest that the design imperatives for the unimon qubit entail (i) close matching of inductive and Josephson energies and (ii) impedance of the order of 1 k$\Omega$. 

\subsection{Qubit coherence optimization}

Aside from studying single-qubit gate infidelity, we have also investigated ways to enhance the coherence times of the unimon qubit by studying $T_{1}$ and $\widetilde{T}_{\varphi}^{1/f}=1/\widetilde{\Gamma}_{\varphi}^{1/f}$ in the region of the parameter space $(E_{\mathrm{J}}/E_{L}, Z)$ which is considered in Sec.~\ref{sec: Gate fidelity optimization}. We have found that $T_{1}$ remains mostly constant when changing $E_{\mathrm{J}}/E_{L}$ and $Z$ at a fixed qubit frequency. However, the dominant mechanism of decoherence in the unimon may vary depending on the choice of parameters. The ratio $\widetilde{T}_{\varphi}^{1/f}/T_{1}$ acts as a useful metric for assessing such changes. If $\widetilde{T}_{\varphi}^{1/f}/T_{1} \gg 1$, the energy relaxation dominates the qubit decoherence, whereas the decoherence is mostly caused by the qubit dephasing if $\widetilde{T}_{\varphi}^{1/f}/T_{1} \ll 1$. 

\begin{figure*}
    \centering
    \includegraphics[width=\linewidth]{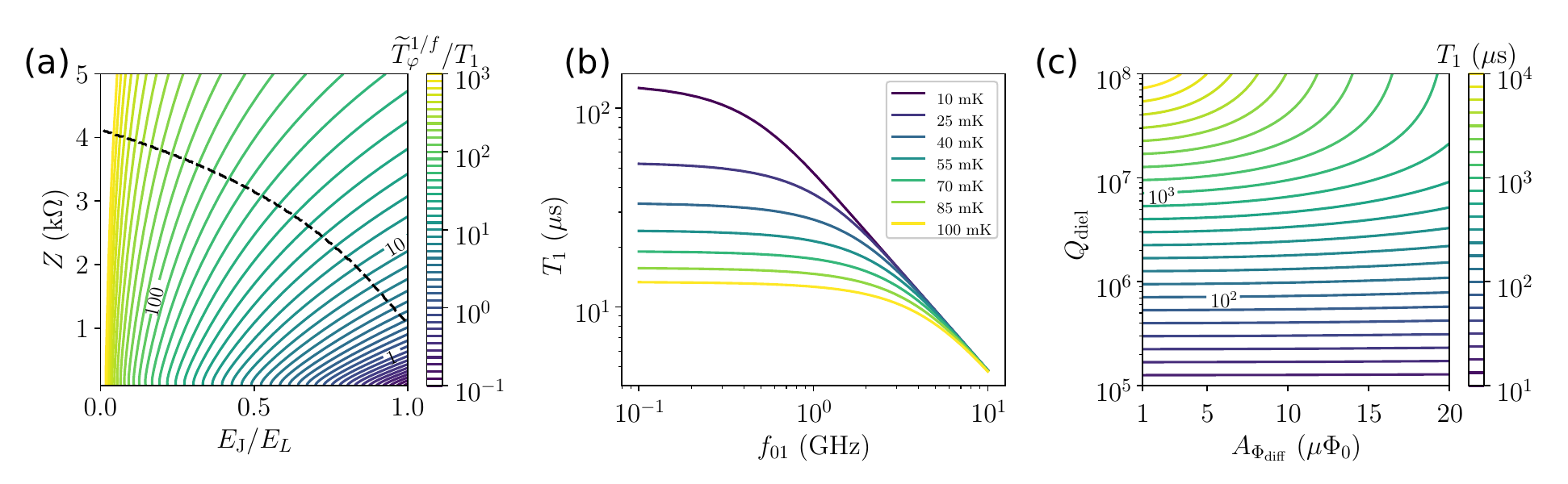}
    \caption{(a) Ratio of the effective pure dephasing time and the energy relaxation time $\widetilde T_{\varphi}^{1/f}/T_{1}$  as a function of the energy ratio $E_{\mathrm{J}}/E_{L}$ and qubit mode impedance $Z$ for $f_{01}=4.5$ GHz, with the dashed line tracing the minimal gate infidelity contour. (b) Relaxation time $T_{1}$ as a function of qubit frequency $f_{01}$ computed for the unimon with parameters $E_{\mathrm{J}}/E_{L}=0.754$ and $Z=315\,\Omega$. Panels (a) and (b) assume $A_{\Phi_{\mathrm{diff}}}$ and $Q_{\mathrm{diel}}$ values provided in Table~\ref{tab:params}. (c) Relaxation time $T_{1}$ as a function of flux noise PSD $A_{\Phi_{\mathrm{diff}}}$ and dielectric quality factor $Q_{\mathrm{diel}}$ computed for the unimon with optimal parameters $E_{\mathrm{J}}/E_{L}=1$, $Z=1\,\mathrm{k}\Omega$, and a qubit frequency $f_{01}=1$ GHz at a temperature $T=25\,\mathrm{mK}$.}
    \label{fig:coherence_optimization}
\end{figure*}

Figure~\ref{fig:coherence_optimization}(a) depicts the ratio $\widetilde{T}_{\varphi}^{1/f}/T_{1}$ across our parametric landscape. We can clearly identify two distinct regions dominated by the different sources of decoherence. In the vicinity of $E_{\mathrm{J}}/E_{L}=0$, the energy relaxation dominates since the  Hamiltonian of the unimon effectively turns into that of a quantum harmonic oscillator. Such a system is completely insensitive to flux noise, but also unusable as a qubit due to the lack of anharmonicity. 
In the vicinity of $E_{\mathrm{J}}/E_{L}=1$ and low impedance, we reach a region dominated by dephasing as a result of an increased flux curvature. Along the trace of minimal gate infidelity as a function of $E_\mathrm{J} /E_L$, the pure dephasing time $\widetilde{T}_{\varphi}^{1/f}$ is roughly an order of magnitude greater than the longitudinal relaxation time $T_{1}$. 

Noting that both $\Gamma_{1}^{1/f}$ and $\Gamma_{1}^{\mathrm{diel}}$ depend on qubit frequency, we study the relaxation time $T_1$  as a function of qubit frequency for fixed $(E_{\mathrm{J}}/E_{L}, Z) = (0.754, 315\,\Omega)$ corresponding to the parameter values of Table~\ref{tab:params}.
As shown in Fig.~\ref{fig:coherence_optimization}(b), the relaxation time of the unimon qubit increases as the qubit frequency decreases. In particular, the relaxation time follows $T_{1} \propto f_{01}^{-1}$ up to the sub-gigahertz frequency range, where temperature effects become a limiting factor for the dielectric loss rate in Eq.~\eqref{eq:gamma_die}. In Sec.~\ref{sec: Gate fidelity optimization}, we observed  that gate fidelity is slightly reduced at low qubit frequencies based on Figs.~\ref{fig:gate_fidelity_optimization}(c)-(f). However, this reduction is only noticeable when $f_{01}$ is of the order of hundreds of MHz, whereas the behavior of infidelity $\bar{\mathcal{E}}$ is comparable across different frequencies down to $f_{01} \sim 1$ GHz in Figs.\ref{fig:gate_fidelity_optimization}(d--f). Consequently, we can achieve a substantial improvement of coherence times without sacrificing fidelity of operation by designing unimon qubits with $f_{01} \approx 1\,\text{GHz}$. This optimization criterion is shared with fluxonia, that operate at subgigahertz qubit frequencies for similar reasons~\cite{Nguyen2019}. Naturally, thermal population is increased for such low qubit frequencies, requiring high-fidelity initialization protocols before computations. 

Let us briefly explore the effects of the flux noise PSD and dielectric quality factor on $T_{1}$. In the calculations above, $A_{\Phi_{\mathrm{diff}}}$ and $Q_{\mathrm{diel}}$ were fixed to values given by the first unimon experiments~\cite{Hyyppa2022}. However, both of these parameters are an order of magnitude lower than in state-of-the-art transmon and fluxonium qubits. Therefore, we carry out a sweep in the $(A_{\Phi_{\mathrm{diff}}}, Q_{\mathrm{diel}})$ space, while setting the unimon qubit parameters to values providing an optimal tradeoff between gate fidelity and energy relaxation time: $f_{01}=1\,\text{GHz}$, $E_{\mathrm{J}}/E_{L}=1$, $Z=1\,\text{k}\Omega$, and $T=25\,\text{mK}$. Figure \ref{fig:coherence_optimization}(c) shows that millisecond coherence times are within reach, provided that the dielectric quality factor can be increased to $\sim10^{7}$ while keeping $A_{\Phi_{\mathrm{diff}}}$ at the previously realized value.

\subsection{Comparison of unimon, transmon and fluxonium}

\begin{figure*}
    \centering
    \includegraphics[width=0.95\linewidth]{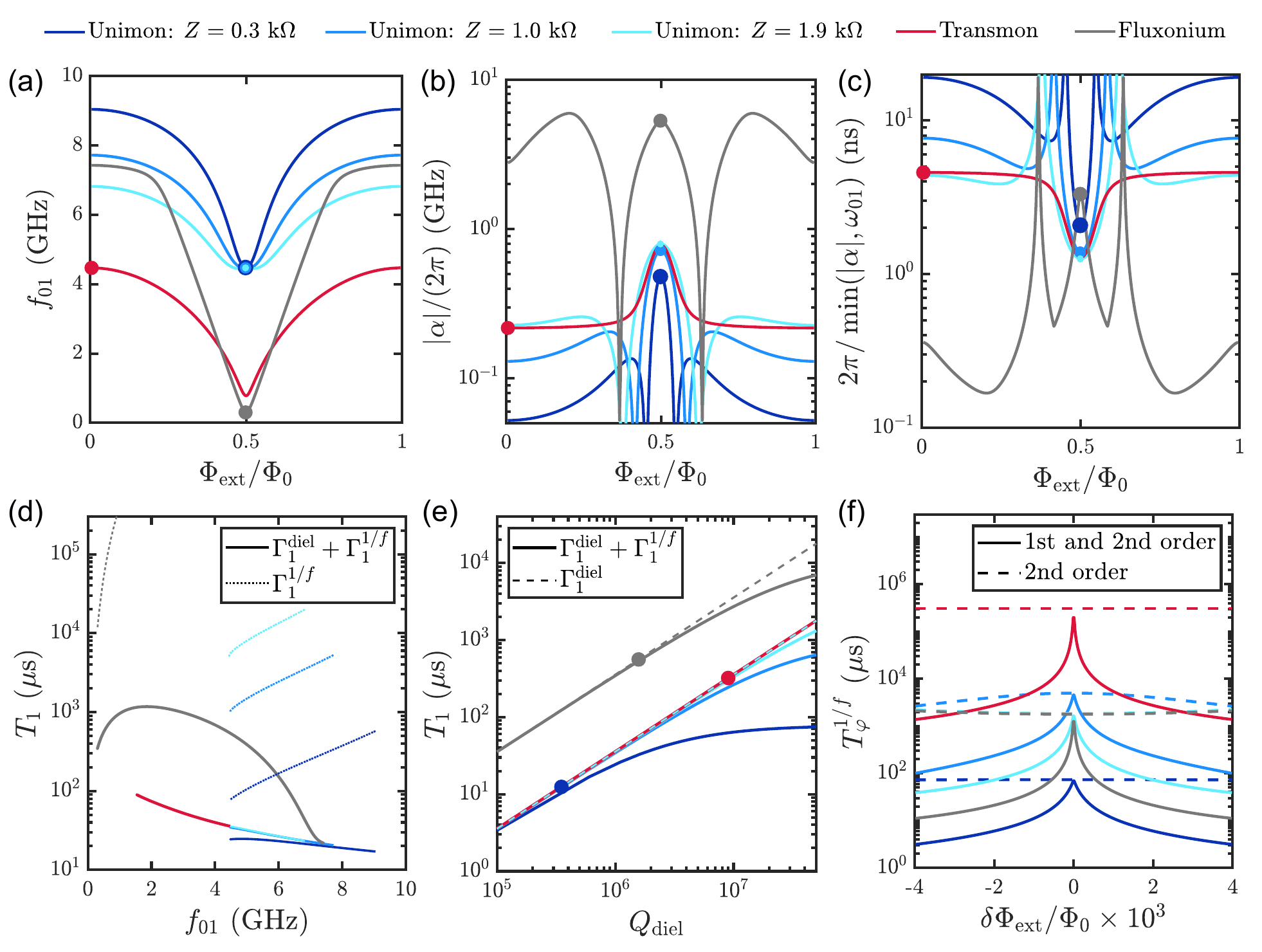}
    \caption{(a)~Qubit frequency $f_{01}$, (b) qubit anharmonicity $\alpha/(2\pi)$, and (c) approximate speed limit of single-qubit gate operation $2\pi/\min(|\alpha|, \omega_{01})$ as functions of the external magnetic flux $\Phi_\mathrm{ext}$ (or $\Phi_\mathrm{diff}$)  for the unimon (different shades of blue as indicated), the transmon (red) and the fluxonium (gray), with the filled circles illustrating the optimal operation point. The qubit parameters are reported in Table~\ref{tab:params for comparison}. For the transmon, we assume an offset charge of $n_\mathrm{g} = 0$.  
    (d) Relaxation time considering both dielectric losses and $1/f$ flux noise as a function of qubit frequency for identical parameters to those used in (a).  For the unimons and the fluxonium, we show the contribution of $1/f$ flux noise (dotted line) based on Eq.~\eqref{eq:gamma_1overf}. For the dielectric loss, we use Eq.~\eqref{eq:gamma_die} and assume a dielectric quality factor of $Q_\mathrm{diel}=10^6$ and a bath temperature of $T=30$~mK for all qubits.  (e) As panel (d) but the relaxation times are shown as functions of the dielectric quality factor. 
    The filled circles illustrate the quality factors achieved in Refs.~\cite{Hyyppa2022,tuokkola2024methods, Ding2023_FTF}. (f) First- and second-order contributions of the $1/f$ flux noise to the dephasing time $T_\varphi^{1/f}$ as functions of the external flux deviation from the optimal operation point of each qubit.
  }
    \label{fig:comparison}
\end{figure*}

The relaxation and coherence times of the first unimon qubits in Ref.~\cite{Hyyppa2022} were lower than those in state-of-the-art transmon and fluxonium qubits which reach up to millisecond-level coherence~\cite{Nguyen2019,Ding2023_FTF,tuokkola2024methods}. 
In Fig.~\ref{fig:comparison}, we numerically compare the frequency, anharmonicity, and coherence properties of unimon, fluxonium, and transmon qubits in order to shed light on the limitations of the first unimon qubits and to demonstrate that an increased mode impedance explored in Sec.~\ref{sec: Gate fidelity optimization} has the potential to significantly improve both the anharmonicity and coherence of the unimon.

In our comparison, we consider three different parameter sets for the unimon that we model using the single-mode Hamiltonian presented in Ref.~\cite{Hyyppa2022}. The first parameter set corresponds to the experimentally characterized values of qubit B in Ref.~\cite{Hyyppa2022} matching parameters of Table~\ref{tab:params}. 
According to the theoretical analysis of Sec.~\ref{sec: Gate fidelity optimization}, the coherence time and gate fidelity of the unimon may be significantly improved by increasing the mode impedance to the kilo-ohm range for $E_\mathrm{J} \lesssim E_L$. Hence, we assume  $Z=1.0$ k$\Omega$ in the parameter set 2 and $Z=1.9$ k$\Omega$ in the parameter set 3, whereas the remaining parameters are chosen to achieve a similar qubit frequency $f_{01} \approx 4.5$ GHz and $E_\mathrm{J}/E_L \approx 0.75$ as in the parameter set 1. Reaching such a
high effective impedance with a CPW structure 
requires a characteristic impedance of $Z_0 \approx 300$ $\Omega$ and $Z_0 \approx 600$ $\Omega$ for the parameter sets 2 and 3, respectively, which is not feasible with geometric inductance alone in a planar CPW geometry. Thus, kinetic inductance \cite{maleeva2018circuit, Grünhaupt2019, Kalacheva2023} or an alternative geometry \cite{peruzzo2021geometric} may be required to reach the required impedance. 
For the transmon and the fluxonium, we select $E_C$, $E_L$, $E_\mathrm{J}$, and the flux noise PSD $A_{\Phi}$ to represent typical values in state-of-the-art devices \cite{Braumuller2020, Nguyen2019, marxer2023long, Ding2023_FTF}. The sweet-spot frequency of the transmon has been chosen to coincide with the corresponding frequency of the unimon. We summarize all the parameters in Table~\ref{tab:params for comparison}.

\begin{table}[t]
    \caption{Parameter values used in the comparison of the unimon, the transmon, and the fluxonium, including  Josephson energy $E_\mathrm{J}$, inductive energy $E_L$, capacitive energy $E_C$, mode impedance $Z$, characteristic impedance $Z_0$ assuming a coplanar waveguide structure of the unimon, flux noise PSD $A_\Phi$ at one hertz, and bath temperature $T$. For the unimon, we consider three sets of parameters, with the first corresponding to the measured values of Qubit B in Ref.~\cite{Hyyppa2022}, and the two other sets representing a higher effective impedance but an equivalent $E_\mathrm{J}/E_L$ ratio and a sweet-spot frequency of $f_{01} \approx 4.5$ GHz at $\Phi_\mathrm{diff}=\Phi_0/2$. We assume the higher effective impedance to be realized with an increased inductance of the center conductor, leading to a reduced length $l$ of the center conductor and a flux noise density scaling as $\sqrt{l}$ \cite{Braumuller2020}.  For the transmon and the fluxonium, the parameters are based on Refs.~\cite{marxer2023long,Ding2023_FTF, Braumuller2020, Nguyen2019}. 
    }
    \begin{center}
        \begin{tabular}{l|l|l|l|l|l|l|l}
            \hline
            \hline
            & $E_{\mathrm{J}}/h$ & $E_{L}/h$ & $E_{C}/h$ & $Z$ & $Z_0$ & $A_{\Phi}$ 
            & $T$ \\
            & (GHz) & (GHz) & (GHz) & (k$\Omega$) & ($\Omega$) & ($\mu \Phi_0$) 
            & (mK)\\ \hline
          Unimon 1 &  19.0 & 25.2 & 0.30 & 0.32 & 97.1 &  15.0 
          & 30 \\
          Unimon 2 &  5.4 & 7.1 & 0.78 & 0.96 &  300 & 9.1 
          & 30  \\
          Unimon 3 &  2.4 & 3.2 & 1.4 & 1.88 & 600 & 6.8 
          & 30 \\
          Transmon &  14.0 & - & 0.195 & - & - & 1.5 
          & 30 \\
          Fluxonium &  6.27 & 0.80 & 1.41 & 3.9 & - & 2.0 
          & 30 \\
            \hline
            \hline
        \end{tabular}
    \end{center}
    \label{tab:params for comparison}
\end{table}

Figures~\ref{fig:comparison}(a) and~\ref{fig:comparison}(b) show the qubit frequency and anharmonicity for the studied qubit parameters. In line with Sec.~\ref{sec: Gate fidelity optimization}, the increased impedance of the unimon both reduces the curvature of the qubit frequency around $\Phi_\textrm{ext} \approx \Phi_0/2$ and increases the anharmonicity to 800~MHz, thus greatly exceeding the typical transmon anharmonicity of around 200--300 MHz. The anharmonicity of the unimon may be further increased by approaching the condition $E_\mathrm{J}/E_L\approx 1$. The fluxonium has an order of magnitude higher anharmonicity compared to the unimon and the transmon, but also an order of magnitude lower frequency. The low qubit frequency may limit the gate speed of the fluxonium to $t_{\mathrm{g}} \sim 1/f_{01}$ when using microwave control techniques \cite{rower2024suppressing, zwanenburg2025single}. Thus, we extend Eq.~\eqref{eq: speed limit} to estimate the speed limit of microwave-based single-qubit gates as $t_{\mathrm{g}, \mathrm{lim}} \sim 2\pi/\min(\omega_{01}, \alpha)$, see Fig.~\ref{fig:comparison}(c). In this comparison, the unimon performs favorably at the optimal operation point compared to the transmon and the fluxonium with the studied parameters. However, the sweet-spot frequency of the fluxonium can in principle be increased by adjusting the device parameters. Note that the bandwidth of the control electronics and pulse distortions may also affect the shortest achievable gate duration in an experimental implementation, in addition to the qubit parameters. Nevertheless, a low value of $t_{\mathrm{g}, \mathrm{lim}}$ is beneficial for preventing issues with frequency crowding. 

Subsequently, we compare the relaxation time of the unimon, the transmon, and the fluxonium in Figs.~\ref{fig:comparison}(d) and~\ref{fig:comparison}(e). Similarly to Sec.~\ref{sec: average gate infidelity estimation}, we consider the relaxation to originate from dielectric losses and $1/f$ flux noise, which are suspected to be the dominant loss mechanisms in the first unimon devices. In Fig.~\ref{fig:comparison}(d), we show the relaxation time owing to $1/f$ flux noise and dielectric losses as a function of qubit frequency. Here, we assume a dielectric quality factor of $Q_\mathrm{diel} = 10^6$ for all the studied qubits in order to compare the protection against dielectric losses arising from the qubit Hamiltonian. For the unimon and the transmon, the relaxation time scales approximately as $T_1 \propto f_{01}^{-1}$ assuming a constant loss tangent \cite{wang2015surface} since the charge matrix elements are approximately equal to those of a quantum harmonic oscillator. In the fluxonium qubit, the relaxation rate owing to dielectric losses is strongly suppressed towards low qubit frequencies due to a reduced charge matrix element \cite{Nguyen2019, somoroff2023millisecond}.  However, the increased thermal noise at low frequencies reduces the relaxation time $T_1 \propto h f_{01}/(2k_\mathrm{B}T)$ around the sweet spot $\Phi_\textrm{ext} = \Phi_0/2$ \cite{Nguyen2019,zhang2021universal, somoroff2023millisecond}, which highlights the importance of thermalization for fluxonium qubits. Nevertheless, the sweet-spot relaxation time caused by dielectric losses may be an order of magnitude lower in fluxonium qubits compared with transmon and unimon qubits with an equivalent dielectric quality factor as shown in Fig.~\ref{fig:comparison}(e).

In the first unimon qubits, the dielectric quality factor was estimated to be $Q_\mathrm{diel} \sim 3 \times 10^5$, which is a factor of 30 lower than in state-of-the-art transmons \cite{tuokkola2024methods} and almost an order of magnitude lower than in state-of-the-art fluxonium qubits \cite{Ding2023_FTF, somoroff2023millisecond}. The fabrication of the first unimon qubits did not utilize state-of-the-art fabrication recipes \cite{tuokkola2024methods, kono2024mechanically}, and hence fabrication improvements have the potential to reduce the dielectric losses. However, a part of the difference in the quality factors may be explained by the interface energy participation ratio of the CPW geometry that is 3-4 times higher than in state-of-the-art parallel plate transmon designs according to FEM simulations~\cite{savola2023design}.

In addition to dielectric losses, we estimate that the $1/f$ flux noise limits the sweet-spot relaxation time of the first unimon qubits to 80~$\mu$s as shown in Fig.~\ref{fig:comparison}(e). An increase of the impedance could drastically enhance the relaxation time owing to flux noise to 1~ms for the parameter set~2 with $Z\approx1$~k$\Omega$ and to 5.6~ms for the parameter set~3 with $Z\approx 1.9$~k$\Omega$. In comparison, the relaxation rate owing to flux noise is around 10~ms at $\Phi_\textrm{ext} = \Phi_0 / 2$ for the studied fluxonium parameters. 

Finally, we show the numerically estimated dephasing time due to the $1/f$ flux noise around the optimal operation point of each qubit. The flux-tunable transmon is best protected against dephasing arising from flux noise due to the small curvature of the qubit frequency around $\Phi_\textrm{ext} = 0$. In contrast to the unimon and the fluxonium, charge noise may limit the dephasing time of transmons below the dephasing time caused by flux noise. Importantly, it seems possible to enhance the sweet-spot dephasing time of the unimon by over an order of magnitude if the impedance is increased to $Z \approx 1 $ k$\Omega$ leading to a lower curvature of the qubit frequency in the vicinity of the sweet spot as shown in Fig.~\ref{fig:comparison}(a). We estimate that the dephasing time of the unimon could significantly exceed the dephasing time of the fluxonium qubit if an increased impedance is used even though the flux noise density were higher for the unimon than for the fluxonium.

\section{Conclusions}
\label{sec:conclusions}
In this work, we studied the quantum-mechanical Hamiltonian of the unimon qubit in order to find conditions for the minimization of the single-qubit gate infidelities. First, we developed closed-form approximations for the qubit frequency and anharmonicity that are generally applicable for single-mode superconducting qubits in the parameter regime $E_{\mathrm{J}} \leq E_{L}$. Our approximations employ the variational method to estimate the lowest qubit eigenenergies with the relative error of the order of $\sim\! 1\%$ in the parameter range relevant for superconducting qubits. The derived equations are simple and accurate, offering a way to estimate the properties of qubits with a single-well potential from the Hamiltonian parameters without having to resort to numerical diagonalization.

We employed these approximations in conjunction with numerical methods to study the gate infidelity over an extended parameter space. We have found that the minimum infidelity is attained at the energy cancellation point $E_{\mathrm{J}}=E_{L}$ with the qubit mode impedance of approximately $1\,\text{k}\Omega$. The location of the minimum is mostly unaffected by the qubit frequency, offering a robust design criterion. Average gate infidelities well below $10^{-4}$ seem attainable provided that the qubit frequency does not exceed the sub-gigahertz domain and that the control electronics do not pose major limitations on the achievable gate duration. 
These results assume $Q_{\mathrm{diel}}$ and $A_{\Phi_{\mathrm{diff}}}$ of the first unimon experiments, suggesting that significant improvements in qubit performance can be achieved even without drastic improvements in the experimental setup or fabrication recipes.

In addition, we investigated ways to improve the coherence times of the unimon qubits. According to our analysis, a low qubit frequency is favorable for the coherence time of the unimon. At sub-gigahertz frequencies, we observe a reduction in gate fidelity, but intermediate qubit frequencies in the vicinity of 1 GHz appear to provide a good balance between long coherence times and high gate fidelities. We have also numerically compared the energy spectra and coherence times of the unimon with its parameters optimized in the above-described way and the state-of-the-art transmon and fluxonium qubits. We conclude that the improved unimon architecture can achieve coherence times comparable to the state-of-the-art transmon qubits with the additional benefit of enhanced anharmonicity. Furthermore, we have found that the dephasing time of the unimon qubit may exceed that of the fluxonium qubits. Clearly, the unimon offers certain advantages 
even when compared to the more popular qubit paradigms. With the parameter optimizations outlined herein, our results pave the way towards the realization of unimon qubits with state-of-the-art qubit properties.

\bibliography{bibliography.bib}

\end{document}